\documentclass[preprint,aps,showpacs]{revtex4}

\usepackage{graphics}

\begin{document}

\title{Self-Adjoint Extensions of the Hamiltonian Operator with
Symmetric Potentials which are Unbounded from Below}

\author{Hing-Tong Cho}
  \email{htcho@mail.tku.edu.tw}
\author{Choon-Lin Ho}
  \email{hcl@mail.tku.edu.tw}
\affiliation{Department of Physics, Tamkang University, Tamsui
251, Taiwan, Republic of China}

\date{May 6, 2008}

\begin{abstract}

We study the self-adjoint extensions of the Hamiltonian operator
with symmetric potentials which go to $-\infty$ faster than
$-|x|^{2p}$ with $p>1$ as $x\rightarrow\pm\infty$. In this
extension procedure, one requires the Wronskian between any states
in the spectrum to approach to the same limit as
$x\rightarrow\pm\infty$. Then the boundary terms cancel and the
Hamiltonian operator can be shown to be hermitian. Discrete bound
states with even and odd parities are obtained. Since the
Wronskian is not required to vanish asymptotically, the energy
eigenstates could be degenerate. Some explicit examples are given
and analyzed.

\end{abstract}

\pacs{03.65.-w, 03.65.Ge, 03.65.Nk, 03.65.Sq}

\maketitle

\section{Introduction}

In quantum mechanics we are used to dealing with potentials that
are bounded from below and are not too singular.  For such
potentials either both bound state wavefunctions and their first
derivatives fall off exponentially at infinities, or the
wavefunctions approach zero with finite first derivatives at the
origin. This in turn implies that for one-dimensional systems the
bound states are non-degenerate as the Wronskians of two
degenerate states vanish. Another ``common sense" in quantum
mechanics is that bound states do not exist for barrier-like
potentials.

Recently, in the course of looking for quasi-exactly solvable
spectra involving quasinormal modes \cite{cho1}, we found a new
quasi-exactly solvable potential,
\begin{equation}
V(x)=-\frac{b^{2}}{4}{\rm
sinh}^{2}x-\left(n^{2}-\frac{1}{4}\right){\rm
sech}^{2}x,\label{QES-potential}
\end{equation}
where $b>0$ is a real parameter and $n=1,2,3,\dots$. Quasi-exactly
solvable models are quantum mechanical systems in which part of
the energy spectrum can be solved exactly
\cite{turbiner,shifman,ushveridze}. The potential in
Eq.~(\ref{QES-potential}) is unbounded from below and is highly
singular at infinities. But interestingly, we found that it admits
quasi-exactly solvable bound states which are doubly-degenerate
(of different parities), and that bound states exist both above
and below the barrier \cite{cho2}.

What we found seems to defy our common sense in quantum mechanics.
However, a careful look at the problem reveals that the
fundamental principles of quantum mechanics are not violated.
Firstly, the Wronskian of any two degenerate states we found does
not vanish.  Secondly, the potential is so singular at infinities
that the time for a classical particle to get to infinity from
some finite reference point $x_{0}$,
\begin{equation}
t_E=\int_{x_{0}}^{\infty}\frac{dt}{\sqrt{2(E-V(x))}}, \label{T}
\end{equation}
is finite.  Hence the unbounded motion in the potential behaves
like as if it were bounded, and the two points at infinities
appear as if they were the end points of a finite segment.
Therefore it appears that bound states are possible and that
degenerate states are allowed.

At first sight, the finiteness of $t_E$ seems to imply that bound
states are possible for any value of $E$.  But this is not true,
as it is not compatible with the requirement that the Hamiltonian
be hermitian.  Recall that for a Hamiltonian $H$ and any generic
wavefunctions $\psi$ and $\phi$, we have
\begin{equation}
\langle\phi|H|\psi\rangle -\langle\psi|H|\phi\rangle \sim \left.
W[\phi,\psi]\right|^{\infty}_{-\infty},\label{sur}
\end{equation}
where $W[\phi,\psi]$ is called the Wronskian  defined by
$W[\phi,\psi]\equiv \phi^\prime \psi-\phi\psi^\prime $ (the prime
means differentiation with respect to the basic variable). The
Hamiltonian $H$ is hermitian if the r.h.s. of Eq.~(\ref{sur})
vanishes. This is usually true for potentials that are bounded
from below and are not too singular. But this is not true for the
potential in Eq.~(\ref{QES-potential}) in general, because
although the wavefunctions vanish at infinities, their first
derivatives diverge at these boundaries. Only for certain discrete
values of energy can $H$ be made hermitian.  Hence boundary
conditions at infinities need be considered in order to extend
Hamiltonians involving singular potentials, such as that in
Eq.~(\ref{QES-potential}), into self-adjoint forms
\cite{neumann,reed}.

Self-adjoint extensions of the Hamiltonian operator with singular
potentials have been discussed before (see e.g.,
\cite{neumann,reed,carreau,feinberg}). These works are concerned
mainly with systems defined on the half-line. On the contrary, the
potential in Eq.~(\ref{QES-potential}) is defined on the whole
line. Thus it seems desirable to give a detailed study of
self-adjoint extension involving symmetric potentials of such a
kind.   This is the main aim of this paper.

This paper is organized as follows. In Sect.~II we discuss parity
states in a generic symmetric barrier-like potential, and
introduce some relevant phases in the WKB wavefunctions. Behaviors
of these phases are analyzed in Sect.~III.  We then introduce in
Sect.~IV the self-adjoint extension procedure. In Sect.~V we
discuss degenerate bound states and total transmission modes with
specific examples.  Sect.~VI concludes the paper.

\section{Parity states}

For symmetric potentials, it is convenient to use states with
definite parity as a basis. To construct them for barrier-like
potentials, we first consider a scattering state, $\psi_{r}$, with
wave coming from the right. Asymptotically, as
$x\rightarrow\infty$, the incident and the reflected waves are
well represented by the WKB wavefunctions,
\begin{equation}
\psi_{r}\sim\frac{1}{\left(E-V(x)\right)^{1/4}}e^{-i\int_{x_{0}}^{x}dx'
\sqrt{E-V(x')}}+\frac{R}{\left(E-V(x)\right)^{1/4}}e^{i\int_{x_{0}}^{x}dx'
\sqrt{E-V(x')}},
\end{equation}
where $R$ is the reflection coefficient and $x_{0}$ is a reference
point. The probability flux can be found to be $-1+|R|^2$. On the
other end, as $x\rightarrow -\infty$,
\begin{equation}
\psi_{r}\sim\frac{T}{\left(E-V(x)\right)^{1/4}}e^{-i\int_{-x_{0}}^{x}dx'
\sqrt{E-V(x')}},
\end{equation}
where $T$ is the transmission coefficient. For symmetric
potentials, we have chosen the reference point on this side to be
$-x_{0}$. The flux here is $-|T|^2$. From flux conservation, we
have
\begin{equation}
|R|^2+|T|^2=1.
\end{equation}
Since the potential is symmetric, one can imagine reversing the
direction to have the wave, $\psi_{l}$, coming from the left with
the same reflection and transmission coefficients. That is,
\begin{eqnarray}
\psi_{l}&\sim&\frac{T}{\left(E-V(x)\right)^{1/4}}e^{i\int_{x_{0}}^{x}dx'
\sqrt{E-V(x')}}\ \ \ \ \ {\rm as}\ x\rightarrow\infty\nonumber\\
\psi_{l}&\sim&\frac{1}{\left(E-V(x)\right)^{1/4}}e^{i\int_{-x_{0}}^{x}dx'
\sqrt{E-V(x')}}+\frac{R}{\left(E-V(x)\right)^{1/4}}e^{-i\int_{-x_{0}}^{x}dx'
\sqrt{E-V(x')}}\nonumber\\ &&\ \ \ \ \ \ \ \ \ \ \ \ \ \ \ \ \ \ \
\ \ \ \ \ \ \ \ \ \ \ \ \ \ \ \ \ \  \ \ \ \ \ \ \ \ \ \ \ \ \ \ \
\ \ \ \ \ \ \ \ \ \ \ \ \ \ \ \ \ \ \ \ \ \ \ {\rm as}\
x\rightarrow -\infty
\end{eqnarray}

Adding these two waves we obtain a state, $\psi^+$, with positive
parity and its asymptotic behavior is
\begin{eqnarray}
\psi^+&\sim&\frac{T+R}{(E-V(x))^{1/4}}e^{i\int_{x_{0}}^{x}dx'\
\sqrt{E-V(x')}}+\frac{1}{(E-V(x))^{1/4}}e^{-i\int_{x_{0}}^{x}dx'\
\sqrt{E-V(x')}}\ \ \ {\rm as}\ x\rightarrow\infty,\nonumber\\
\psi^+&\sim&\frac{1}{(E-V(x))^{1/4}}e^{i\int_{-x_{0}}^{x}dx'\
\sqrt{E-V(x')}}+\frac{T+R}{(E-V(x))^{1/4}}e^{-i\int_{-x_{0}}^{x}dx'\
\sqrt{E-V(x')}}\ \ {\rm as}\ x\rightarrow-\infty.\nonumber\\
\end{eqnarray}
Similarly, subtracting the two waves we obtain the negative parity
state, $\psi^-$, with the asymptotic behavior,
\begin{eqnarray}
\psi^-&\sim&\frac{T-R}{(E-V(x))^{1/4}}e^{i\int_{x_{0}}^{x}dx'\
\sqrt{E-V(x')}}-\frac{1}{(E-V(x))^{1/4}}e^{-i\int_{x_{0}}^{x}dx'\
\sqrt{E-V(x')}}\ \ \ {\rm as}\ x\rightarrow\infty,\nonumber\\
\psi^-&\sim&\frac{1}{(E-V(x))^{1/4}}e^{i\int_{-x_{0}}^{x}dx'\
\sqrt{E-V(x')}}-\frac{T-R}{(E-V(x))^{1/4}}e^{-i\int_{-x_{0}}^{x}dx'\
\sqrt{E-V(x')}}\ \ {\rm as}\ x\rightarrow-\infty.\nonumber\\
\end{eqnarray}

Since the fluxes of these parity states are zero, we must have
\begin{equation}
|T\pm R|^{2}-1=0~~~\Rightarrow ~~T^{*}R+TR^{*}=0,
\end{equation}
that is, $T^{*}R$ is purely imaginary. Hence, one can parametrize
$T$ and $R$ as
\begin{equation}
T=\cos\alpha\ e^{i\theta}\ \ ,\ \ R=-i\sin\alpha\
e^{i\theta},\label{alphatheta}
\end{equation}
for $0\leq \alpha\leq \pi/2$. Here we have incorporated the fact
that $|T|^2+|R|^2=1$ into the parametrization. Note that
$\cos^{2}\alpha$ is just the transmission probability. Above the
barrier, $\alpha\sim 0$, while $\alpha\sim\pi/2$ for tunneling
through the barrier. Furthermore, we have
\begin{equation}
T\pm R=\left(\cos\alpha\mp
i\sin\alpha\right)e^{i\theta}=e^{i\left(\theta\mp \alpha\right)}.
\end{equation}
In terms of these phases, $\alpha$ and $\theta$, the asymptotic
forms of $\psi^+$ and $\psi^-$ can be rewritten in an explicitly
flux-zero real form. For $\psi^+$,
\begin{eqnarray}
\psi^+&\sim&\frac{1}{(E-V(x))^{1/4}}\cos\left[
\int_{x_{0}}^{x}dx'\
\sqrt{E-V(x')}+\frac{1}{2}\left(\theta-\alpha\right)\right]\ \ \ \
\ {\rm as}\ x\rightarrow\infty,\nonumber\\
\psi^+&\sim&\frac{1}{(E-V(x))^{1/4}}\cos\left[
\int_{-x_{0}}^{x}dx'\
\sqrt{E-V(x')}-\frac{1}{2}\left(\theta-\alpha\right)\right]\ \ \ \
\ {\rm as}\ x\rightarrow-\infty,\label{psi1}
\end{eqnarray}
and for $\psi^-$,
\begin{eqnarray}
\psi^-&\sim&\frac{1}{(E-V(x))^{1/4}}\sin\left[
\int_{x_{0}}^{x}dx'\
\sqrt{E-V(x')}+\frac{1}{2}\left(\theta+\alpha\right)\right]\ \ \ \
\ {\rm as}\ x\rightarrow\infty,\nonumber\\
\psi^-&\sim&\frac{1}{(E-V(x))^{1/4}}\sin\left[
\int_{-x_{0}}^{x}dx'\
\sqrt{E-V(x')}-\frac{1}{2}\left(\theta+\alpha\right)\right]\ \ \ \
\ {\rm as}\ x\rightarrow-\infty.\label{psi2}
\end{eqnarray}
Note that up to here, our construction is completely general.
These parity states can be defined for any symmetric barrier-like
potentials. On the other hand, for the potentials we shall
consider in the following sections, these states are normalizable
due to the $(E-V)^{-1/4}$ factor, and under the self-adjoint
extension procedure they become possible bound states for the
system.


\section{The behavior of various phases}

In the following we concentrate on symmetric potentials which go
to minus infinity asymptotically. If the rate is faster than
$|x|^{2p}$, with $p>1$, then from Eq.~(\ref{T}) the time for a
classical particle to get to infinity from some finite reference
point $x_{0}$ is finite. Quantum mechanically a wave packet will
get to infinity in finite time and then the probability is lost.
In order to have a hermitian Hamiltonian, self-adjoint extensions
must be defined.

To make sure that no probability is lost at infinity, we use the
parity states defined in the last section as a basis. They are
explicitly flux-zero with the asymptotic behavior given by Eqs.
(\ref{psi1}) and (\ref{psi2}). Before we analyze these asymptotic
expressions in more details, we need to define yet another phase,
the WKB phase $\phi$. Consider in Eqs. (\ref{psi1}) and
(\ref{psi2}) the integral,
\begin{eqnarray}
\int_{x_{0}}^{x}dx'\ \sqrt{E-V(x')}&=&\int_{x_{0}}^{\infty}dx\
\sqrt{E-V(x)}-\int_{x}^{\infty}dx'\ \sqrt{E-V(x')},
\end{eqnarray}
Since $x$ is large, $|V|\gg E$, one can expand the integrand in
the second integral,
\begin{eqnarray}
\int_{x}^{\infty}dx'\ \sqrt{E-V(x')}&=&\int_{x}^{\infty}dx'\
\sqrt{-V(x')}\left(1-\frac{E}{V(x')}\right)^{1/2}\nonumber\\
&=&\int_{x}^{\infty}dx'\ \sqrt{-V(x')}+\cdots\nonumber\\
&=&-\int_{0}^{x}dx'\sqrt{-V(x')}+\int_{0}^{\infty}dx\sqrt{-V(x)}+\cdots,
\end{eqnarray}
then
\begin{eqnarray}
&&\int_{x_{0}}^{x}dx'\ \sqrt{E-V(x')}\nonumber\\
&=&\int_{0}^{x}dx'\ \sqrt{-V(x')}+\int_{x_{0}}^{\infty}dx
\left(\sqrt{E-V(x)}-\sqrt{-V(x)}\right)-\int_{0}^{x_{0}}dx\sqrt{-V(x)}
+\cdots,
\end{eqnarray}
with the remaining part vanishing as $x\rightarrow\infty$. Note
that here we have assumed that, without loss of generality, the
maximum value of the potential is $V=0$. Next, we define the WKB
phase,
\begin{equation}
\phi\equiv\int_{x_{0}}^{\infty}dx
\left(\sqrt{E-V(x)}-\sqrt{-V(x)}\right)-\int_{0}^{x_{0}}dx\sqrt{-V(x)},
\end{equation}
which is finite for a fixed value of $E$ for the potentials that
we are considering here. Then the asymptotic behavior for the
parity states can be rewritten as
\begin{eqnarray}
\psi^+&\sim&\frac{1}{(-V(x))^{1/4}}\cos\left[ \int_{0}^{x}dx'\
\sqrt{-V(x')}+\phi+\frac{1}{2}\left(\theta-\alpha\right)\right]\ \
\ \ \ {\rm as}\ x\rightarrow\infty,\nonumber\\
\psi^+&\sim&\frac{1}{(-V(x))^{1/4}}\cos\left[ \int_{0}^{x}dx'\
\sqrt{-V(x')}-\phi-\frac{1}{2}\left(\theta-\alpha\right)\right]\ \
\ \ \ {\rm as}\ x\rightarrow-\infty,\label{psi11}
\end{eqnarray}
and
\begin{eqnarray}
\psi^-&\sim&\frac{1}{(-V(x))^{1/4}}\sin\left[ \int_{0}^{x}dx'\
\sqrt{-V(x')}+\phi+\frac{1}{2}\left(\theta+\alpha\right)\right]\ \
\ \ \ {\rm as}\ x\rightarrow\infty,\nonumber\\
\psi^-&\sim&\frac{1}{(-V(x))^{1/4}}\sin\left[ \int_{0}^{x}dx'\
\sqrt{-V(x')}-\phi -\frac{1}{2}\left(\theta+\alpha\right)\right]\
\ \ \ \ {\rm as}\ x\rightarrow-\infty.\label{psi12}
\end{eqnarray}


Now we are ready to analyze the behaviors of various phases
$\phi$, $\theta$, and $\alpha$ as functions of the energy $E$. We
shall use the potential $V(x)=-a^{2}|x|^{2p}$ as a concrete
example when necessary to elucidate the general arguments. Some
more examples will be discussed in Section~V.

First, we look at the WKB phase $\phi$. This phase vanishes at
$E=0$ when the reference point $x_{0}$ is chosen to be the origin.
For other values of $E$, it increases monotonically from negative
infinity to positive infinity as $E$ is increased.

For example, for the potential $V(x)=-a^{2}|x|^{2p}$, $p>1$, we
can evaluate the WKB phase explicitly. For $E<0$, we shall choose
the reference point to be the classical turning point
$x_{0}=a^{-1/p}(-E)^{1/2p}$. In this case the WKB phase becomes
\begin{eqnarray}
\phi(E)&=&\int_{x_{0}}^{\infty}dx\left(\sqrt{E+a^{2}x^{2p}}-ax^{p}\right)
-\int_{0}^{x_{0}}dx\ ax^{p}\nonumber\\
&=&-a^{-\frac{1}{p}}(-E)^{\frac{1}{2}\left(\frac{1}{p}+1\right)}A(p),\label{wkbphase1}
\end{eqnarray}
where the $p$-dependent positive constant
\begin{equation}
A(p)=\int_{1}^{\infty}d\xi(\xi^{p}-\sqrt{\xi^{2p}-1})+\int_{0}^{1}d\xi\
\!\xi^{p}
=\frac{p\sqrt{\pi}\Gamma\left(\frac{3}{2}-\frac{1}{2p}\right)}
{(p-1)(p+1)\Gamma\left(1-\frac{1}{2p}\right)},
\end{equation}
with $A(2)\sim 0.87$. Here the WKB phase decreases as
$-(-E)^{\frac{1}{2}\left(\frac{1}{p}+1\right)}$ as $E$ is
decreased to minus infinity.

For $E>0$, one can choose the reference point $x_{0}$ to be at the
origin. Hence,
\begin{eqnarray}
\phi(E)
&=&a^{-\frac{1}{p}}E^{\frac{1}{2}\left(\frac{1}{p}+1\right)}B(p),\label{wkbphase2}
\end{eqnarray}
where
\begin{equation}
B(p)=\int_{0}^{\infty}d\zeta\left(\sqrt{1+\zeta^{2p}}-\zeta^{p}\right)
=
\frac{p\Gamma\left(\frac{1}{2p}\right)\Gamma\left(\frac{3}{2}-\frac{1}{2p}\right)}
{(p-1)(p+1)\sqrt{\pi}},\label{defb}
\end{equation}
is a $p$-dependent positive constant. For example, $B(2)\sim
1.23$. We can see that as $E$ increases, the WKB phase also
increases as $E^{\frac{1}{2}(\frac{1}{p}+1)}$. In summary, the WKB
phase increases monotonically as
$|E|^{\frac{1}{2}\left(\frac{1}{p}+1\right)}$ from minus infinity
to plus infinity as $E$ is increased.

The phases $\alpha$ and $\theta$ are related to the transmission
and the reflection coefficients in the scattering process as given
in Eq.~(\ref{alphatheta}). We know that $0\leq\alpha\leq\pi/2$.
For $\theta$ we shall see in the following that it is always
small.

For $E<0$, one can estimate the transmission coefficient $T$ using
the usual WKB approximation. To the lowest order, we have
\begin{equation}
T\sim e^{-2\beta},
\end{equation}
where $\beta=\int_{0}^{x_{0}}dx\sqrt{V(x)-E}$. For the potential
$V(x)=-a^{2}|x|^{2p}$,
$\beta=a^{-\frac{1}{p}}(-E)^{\frac{1}{2}\left(\frac{1}{p}+1\right)}C(p)$
with the $p$-dependent constant
\begin{equation}
C(p)=\int_{0}^{1}d\xi\sqrt{1-\xi^{2p}}
=\frac{\sqrt{\pi}\Gamma\left(1+\frac{1}{2p}\right)}
{2\Gamma\left(\frac{3}{2}+\frac{1}{2p}\right)}.\label{defc}
\end{equation}
It is interesting to see that this constant $C(p)$ is related to
the constant $B(p)$ in Eq.~(\ref{defb}) by
\begin{equation}
C(p)=B(p)\cos\left(\frac{\pi}{2p}\right).\label{cbrelation}
\end{equation}
This relation will be important when we consider the condition for
total transmission modes later.
 Hence, for $E<0$, the
phases
\begin{equation}
\alpha\sim\cos^{-1}\left(e^{-2\beta}\right)\sim\frac{\pi}{2}-e^{-2\beta}\
\ ,\ \ \theta\sim 0.
\end{equation}

For $E>0$, the estimation of the phases is more subtle because
there are no turning points. A direct application of the WKB
formula will always give no reflection. To deal with this
situation, one can extend the WKB approximation by using complex
contour method (see, e.g., \cite{berry}). Then a formula for the
reflection coefficient above the barrier can be obtained related
to the complex turning points.
\begin{equation}
R\sim\sum_{j}\left(-\frac{i\pi}{3}\right)e^{2i\gamma_{j}},
\end{equation}
where
\begin{equation}
\gamma_{j}=\int_{0}^{x_{j}}dx\sqrt{E-V(x)},\label{gamma}
\end{equation}
and $x_{j}$ is a complex turning point, or a complex root of
$E-V(x)=0$ in the upper half $x$-plane. The main contribution to
$R$ comes from the roots nearest to the real axis. For the
potential, $V(x)=-a^{2}|x|^{2p}$, the two roots nearest to the
real axis are $x_{1}=(E/a^{2})^{1/2p}e^{i\pi/2p}$ and
$x_{2}=(E/a^{2})^{1/2p}e^{i\pi(1-1/2p)}$. The corresponding
$\gamma$'s are
\begin{eqnarray}
\gamma_{1}&=&\int_{0}^{x_{1}}dx\sqrt{E+a^{2}x^{2p}}
=a^{-\frac{1}{p}}E^{\frac{1}{2}\left(\frac{1}{p}+1\right)}
e^{i\frac{\pi}{2p}}C(p),\nonumber\\
\gamma_{2}&=&-a^{-\frac{1}{p}}E^{\frac{1}{2}\left(\frac{1}{p}+1\right)}
e^{-i\frac{\pi}{2p}}C(p).
\end{eqnarray}
In this case, the reflection coefficient can be estimated to be
\begin{eqnarray}
R\sim-i\left(\frac{2\pi}{3}\right)\cos\left[2a^{-\frac{1}{p}}
E^{\frac{1}{2}\left(\frac{1}{p}+1\right)}C(p)\cos\left(\frac{\pi}{2p}\right)\right]
e^{-2a^{-\frac{1}{p}}
E^{\frac{1}{2}\left(\frac{1}{p}+1\right)}C(p)\sin\left(\frac{\pi}{2p}\right)}.
\end{eqnarray}
Therefore, for $E>0$, the phases $\alpha$ and $\theta$ are
\begin{eqnarray}
\alpha&\sim&\left(\frac{2\pi}{3}\right)\cos\left[2a^{-\frac{1}{p}}
E^{\frac{1}{2}\left(\frac{1}{p}+1\right)}C(p)\cos\left(\frac{\pi}{2p}\right)\right]
e^{-2a^{-\frac{1}{p}}
E^{\frac{1}{2}\left(\frac{1}{p}+1\right)}C(p)\sin\left(\frac{\pi}{2p}\right)},\nonumber\\
\theta&\sim&0.\label{ttcondition}
\end{eqnarray}

By these WKB estimates, we see that the phase $\theta\sim 0$ in
the whole energy range. For the phase $\alpha$, it is almost equal
to $\pi/2$ for $E<0$, and then it decreases to exponentially small
values for $E>0$ as given by Eq.~(\ref{ttcondition}). Hence, in
general the phase $\theta$ is small for all values of $E$, while
the phase $\alpha$ changes from $\pi/2$ (for
$E\rightarrow-\infty$) to 0 (for $E\rightarrow\infty$), with
possible oscillations in which $R\sim 0$.

From the expression in Eq.~(\ref{ttcondition}), one can obtain the
condition for $R\sim 0$, that is, the energies at which total
transmission occurs. Here we have
\begin{eqnarray}
&&2a^{-\frac{1}{p}}
E_{n}^{\frac{1}{2}\left(\frac{1}{p}+1\right)}C(p)\cos\left(\frac{\pi}{2p}\right)
=(2n+1)\frac{\pi}{2}\nonumber\\
&\Rightarrow&E_{n}=\left[\frac{(2n+1)\pi a^{\frac{1}{p}}}
{4C(p)\cos\left(\frac{\pi}{2p}\right)}\right]^{\frac{2p}{p+1}}.\label{x2ptt}
\end{eqnarray}
For example, for $a=1$ and $p=2$,
\begin{equation}
E_{0}=1.3765,\ \ E_{1}=5.9558,\ \ E_{2}=11.769,\ \ \dots
\end{equation}
The total transmission modes occur for all values of $p>1$. It is
therefore a generic phenomenon that they are present for symmetric
potentials unbounded from below.

\section{Self-adjoint extension}

The parity states with the asymptotic behaviors in
Eqs.~(\ref{psi11}) and (\ref{psi12}) are normalizable because of
the $1/(-V)^{1/4}$ factor. However, just because of this, as we
have discussed earlier, a wavepacket will get to infinity in
finite time and the probability is lost. This manifests itself in
the fact that the Wronskian of two wavefunctions (assumed to be
real)
\begin{equation}
W[\psi_{1},\psi_{2}]\equiv\frac{d\psi_{1}}{dx}\psi_{2}-
\psi_{1}\frac{d\psi_{2}}{dx},
\end{equation}
may not necessarily vanish asymptotically as
$x\rightarrow\pm\infty$. Hence, self-adjoint extensions have to be
implemented to have a hermitian Hamiltonian operator. Here for the
symmetric potential case we are considering, one only needs to
require that the Wronskian between any two states goes to the same
limit as $x\rightarrow\pm\infty$. Then the boundary terms cancel
in the integral and the Hamiltonian operator can be shown to be
hermitian.


To implement the self-adjoint extension procedure, we choose a
positive parity reference state $\psi^+_{0}$ with energy
$E^{+}_{0}$. From Eq.~(\ref{psi11}), we see that as
$x\rightarrow\infty$,
\begin{equation}
\psi^+_{0}\sim\frac{1}{(-V(x))^{1/4}}\cos\left[ \int_{0}^{x}dx'\
\sqrt{-V(x')}+\phi^+_{0}+\frac{1}{2}\left(\theta^+_{0}-\alpha^+_{0}\right)\right].
\end{equation}
For another positive parity state $\psi^+_{n}$, the asymptotic
behavior of the Wronskian
\begin{eqnarray}
\left[\frac{d\psi^+_0}{dx}\psi^+_n-\psi^+_0\frac{d\psi^+_n}{dx}\right]
_{x\rightarrow\infty}
&=&-\sin\left[\int_{0}^{x}dx'\sqrt{-V(x')}+\phi^+_0+\frac{1}{2}
\left(\theta^+_0-\alpha^+_0\right)\right]\nonumber\\ &&\ \ \ \
\times\cos\left[\int_{0}^{x}dx'\sqrt{-V(x')}+\phi^{+}_n+\frac{1}{2}
\left(\theta^+_n-\alpha^+_n\right)\right]\nonumber\\ && +
\cos\left[\int_{0}^{x}dx'\sqrt{-V(x')}+\phi^+_0+\frac{1}{2}
\left(\theta^+_0-\alpha^+_0\right)\right]\nonumber\\ &&\ \ \ \
\times\sin\left[\int_{0}^{x}dx'\sqrt{-V(x')}+\phi^{+}_n+\frac{1}{2}
\left(\theta^+_n-\alpha^+_n\right)\right]\nonumber\\
&=&\sin\left[\left(\phi^{+}_n+\frac{1}{2}\left(\theta^+_n-\alpha^+_n\right)\right)
-\left(\phi^+_0+\frac{1}{2}\left(\theta^+_0-\alpha^+_0\right)\right)\right].\nonumber\\
\label{wronasym}
\end{eqnarray}
Since the Wronskian of two positive parity states is odd, one must
require this limit to vanish to have the same limit as
$x\rightarrow\pm\infty$. Therefore, the two asymptotic phases in
Eq.~(\ref{wronasym}) must be the same modulo $n\pi$, that is,
\begin{equation}
\phi^{+}_n+\frac{1}{2}\left(\theta^+_n-\alpha^+_n\right)
=\phi^+_0+\frac{1}{2}\left(\theta^+_0-\alpha^+_0\right)+n\pi,\label{positive}
\end{equation}
where $n=0,\pm 1,\pm 2,\dots$. In the discussion of the preceding
section, we know that the WKB phase $\phi$, as indicated in
Eqs.~(\ref{wkbphase1}) and (\ref{wkbphase2}), increases with the
energy $E$, while the other two phases are small. Hence, there are
infinite number of energies (or states) which satisfy the relation
in Eq.~(\ref{positive}). This is the quantization rule for the
positive parity part of the discrete set of bound states in this
self-adjoint extension.

Similarly, one considers a set of negative parity states
$\psi^{-}_{m}$. Take $\psi_{0}^{-}$ as the negative parity
reference state with energy $E_{0}^{-}$. From the asymptotic
behavior of negative parity states as shown in Eq.~(\ref{psi12}),
we have, as $x\rightarrow\infty$,
\begin{eqnarray}
\psi^-_{m}&\sim&\frac{1}{(-V(x))^{1/4}}\sin\left[ \int_{0}^{x}dx'\
\sqrt{-V(x')}+\phi^{-}_{m}+\frac{1}{2}\left(\theta^{-}_{m}+\alpha^{-}_{m}\right)\right].
\end{eqnarray}
Since the Wronskian of two negative parity states is again odd, we
must require the Wronskian to vanish as $x\rightarrow\pm\infty$ as
before. Then we obtain the quantization rule for the negative
parity part of the spectrum as given by
\begin{equation}
\phi^{-}_m+\frac{1}{2}\left(\theta^-_m+\alpha^-_m\right)=
\phi^-_0+\frac{1}{2}\left(\theta^-_0+\alpha^-_0\right)+m\pi,
\label{2schemen}
\end{equation}
where $m=0,\pm 1,\pm 2,\dots$.

The Wronskian of a positive parity state and a negative parity
state is always even so it will approach to the same limit as
$x\rightarrow\pm\infty$. Therefore, a set of positive parity
states satisfying Eq.~(\ref{positive}) can combine with any set of
negative parity states satisfying Eq.~(\ref{2schemen}) to give a
discrete bound state spectrum. Since $E^+_0$ and $E^-_0$ are
arbitrary, we have a two-parameter family of spectra. This
constitutes the self-adjoint extension procedure for the
Hamiltonian operator to be hermitian.

There are some special cases one can consider. For example, when
the reference state energies $E_{0}^{+}$ and $E_{0}^{-}$ are the
same, the spectrum is characterized by only one free parameter. In
this case, the reference states constitute a pair of degenerate
states. Actually, it is possible to have many pairs of degenerate
states which are total transmission or reflectionless. This
possibility will be explored in some details in the next section.

Another interesting special case is that the Wronskians between
any states in the spectrum vanish asymptotically as
$x\rightarrow\pm\infty$. This is already true for states with the
same parity. To examine the Wronskians between opposite parity
states, we take, for example the reference state $\psi_{0}^{+}$
and any state $\psi_{m}^{-}$ in the negative parity set. The
Wronskian between $\psi^+_0$ and $\psi^-_m$ has the asymptotic
behavior,
\begin{eqnarray}
&&\left[\frac{d\psi^+_0}{dx}\psi^-_m-\psi^+_0\frac{d\psi^-_m}{dx}\right]
_{x\rightarrow\infty}\nonumber\\
&=&-\sin\left[\int_{0}^{x}dx'\sqrt{-V(x')}+\theta^+_0+\frac{1}{2}
\left(\theta^+_0-\alpha^+_0\right)\right]\nonumber\\ &&\ \ \ \
\times\sin\left[\int_{0}^{x}dx'\sqrt{-V(x')}+\phi^{-}_m+\frac{1}{2}
\left(\theta^-_m+\alpha^-_m\right)\right]\nonumber\\ && -
\cos\left[\int_{0}^{x}dx'\sqrt{-V(x')}+\phi^+_0+\frac{1}{2}
\left(\theta^+_0-\alpha^+_0\right)\right]\nonumber\\ &&\ \ \ \
\times\cos\left[\int_{0}^{x}dx'\sqrt{-V(x')}+\phi^{-}_m+\frac{1}{2}
\left(\theta^-_m+\alpha^-_m\right)\right]\nonumber\\
&=&-\cos\left[\left(\phi^{-}_m+\frac{1}{2}\left(\theta^-_m+\alpha^-_m\right)\right)
-\left(\phi^+_0+\frac{1}{2}\left(\theta^+_0-\alpha^+_0\right)\right)\right].
\nonumber\\
\end{eqnarray}
For this limit to vanish, we must have the requirement,
\begin{equation}
\phi^{-}_m+\frac{1}{2}\left(\theta^-_m+\alpha^-_m\right)=
\phi^+_0+\frac{1}{2}\left(\theta^+_0-\alpha^+_0\right)+\left(m+\frac{1}{2}\right)\pi,
\label{positivenegative}
\end{equation}
where $m=0,\pm 1,\pm 2,\dots$. From the asymptotic behavior of the
positive parity states in Eq.~(\ref{positive}), we can see that
any negative parity state will have asymptotically vanishing
Wronskians with all the positive parity states. This set of
positive and negative parity states is parametrized only by one
free parameter, the energy $E_{0}^{+}$ of the reference state
$\psi^+_0$. One could of course start with any positive or
negative parity state in the set as the reference state. The same
Hilbert space in which the Hamiltonian operator is hermitian will
be obtained.

\section{Degenerate and total transmission states}

In the preceding section, the self-adjoint extension of the
Hamiltonian operator with symmetric potentials which are unbounded
from below are introduced. In this procedure, since the Wronskian
is not required to vanish at infinity, the energy eigenstates
could be degenerate. Let us elaborate on this point in more
details. Suppose one chooses the reference states in the positive
and negative parity sector to be of the same energy. Then the
quantization condition on the positive parity sector is given by
\begin{equation}
\phi^{+}_n+\frac{1}{2}\left(\theta^+_n-\alpha^+_n\right)=\phi_{0}
+\frac{1}{2}\left(\theta_{0}-\alpha_{0}\right)+n\pi,
\label{sameref1}
\end{equation}
and that on the negative parity section is given by
\begin{equation}
\phi^{-}_m+\frac{1}{2}\left(\theta^-_m+\alpha^-_m\right) =\phi_{0}
+\frac{1}{2}\left(\theta_{0}+\alpha_{0}\right)+m\pi.
\label{sameref2}
\end{equation}
We drop the parity signs on the r.h.s. of Eqs.~(\ref{sameref1})
and (\ref{sameref2}) since $\phi^+_{0}=\phi^-_{0}=\phi_0$ and so
on. Although the reference states are degenerate, the other states
in the positive and the negative parity sectors may not be because
the difference in the quantization in Eqs.~(\ref{sameref1}) and
(\ref{sameref2}) of the two sectors.

A special but very interesting situation occurs when the reference
states are total transmission modes, that is $\alpha_{0}=0$. In
addition to choosing these reference states, if the condition for
total transmission coincides with the quantization rule in the
self-adjoint extension requirement, we shall have
\begin{equation}
\phi^\pm_n+\frac{1}{2}\theta^\pm_n=\phi_{0}+\frac{1}{2}\theta_{0}+n\pi,\label{ttselfadj}
\end{equation}
for $n=0,1,2,\dots$. Then the positive and the negative parity
sectors have the same quantization rule, and the total
transmission states are all degenerate.

These criteria seem to be very stringent but there are potentials
that satisfy them. Going back to the potential
$V(x)=-a^{2}|x|^{2p}$ we have been considering so far. For $E>0$,
the phases are given by Eqs.~(\ref{wkbphase2}) and
(\ref{ttcondition}). The occurrence of the total transmission
modes for $\alpha=0$ and $\theta=0$ is at energies given by
Eq.~(\ref{x2ptt}). From this condition, we can write the energies
as
\begin{equation}
E_{n}^{\frac{p+1}{2p}}-E_{0}^{\frac{p+1}{2p}}=\frac{n\pi
a^{\frac{1}{p}}}{2C(p)\cos\frac{\pi}{2p}},\label{x4ttenergy}
\end{equation}
for $n=0,1,2,\dots$, where we have taken the lowest energy total
transmission mode as the reference. Next, we calculate the WKB
phases of these total transmission modes. From
Eqs.~(\ref{wkbphase2}), (\ref{cbrelation}), and
(\ref{x4ttenergy}),
\begin{equation}
\phi_{n}=a^{-\frac{1}{p}}E_{n}^{\frac{p+1}{2p}}B(p)
~~~~\Rightarrow~~\phi_{n}-\phi_{0}=\frac{n\pi}{2\cos^{2}\left(\frac{\pi}{2p}\right)}.
\end{equation}
This coincides with the quantization rule from the self-adjoint
extension requirement if
\begin{equation}
\frac{1}{2\cos^{2}\left(\frac{\pi}{2p}\right)}=1~~~~\Rightarrow
~~p=2.
\end{equation}
Hence, for the potential $V(x)=-a^{2}x^{4}$, if we choose the
reference states to be the positive and the negative parity states
of a total transmission mode, the states with $E>0$ in the
spectrum are all doubly degenerate. Note that the part of the
spectrum with $E<0$ is not degenerate because $\alpha$ is nonzero
there.

There is yet another potential which satisfies these criteria,
namely, the potential in Eq.~(\ref{QES-potential}) we mentioned in
the Introduction.  This potential is in fact quasi-exactly
solvable. For this potential, there are $n$ states which are
exactly solvable. They are found to be total transmission states.
For example, for $n=2$, we have the solvable right ($r$) and left
($l$) moving total transmission modes,
\begin{equation}
\psi_{1(r,l)}=\frac{e^{\pm\frac{i}{2}b{\rm sinh}x}}{({\rm cosh}
x)^{3/2}}\left[\pm i\ \!{\rm sinh}
x-\frac{1}{b}\left(\sqrt{b^{2}+1}+1\right)\right],
\end{equation}
with energy $E_{1}=\frac{1}{4}(b^{2}-5)-\sqrt{b^{2}+1}$, and
\begin{equation}
\psi_{2(r,l)}=\frac{e^{\pm\frac{i}{2}b{\rm sinh}x}}{({\rm cosh}
x)^{3/2}}\left[\pm i\ \!{\rm sinh}
x-\frac{1}{b}\left(\sqrt{b^{2}+1}-1\right)\right],
\end{equation}
with energy $E_{2}=\frac{1}{4}(b^{2}-5)+\sqrt{b^{2}+1}$.  From
these total transmission modes, one can construct the positive and
the negative parity states,
\begin{eqnarray}
\psi^+_{(1,2)}&=&\left({\rm cosh}x\right)^{-3/2}\left[
\cos\left(\frac{b}{2}{\rm
sinh}x\right)-\frac{1}{b}(1\mp\sqrt{b^{2}+1}){\rm sinh}x\
\!\sin\left(\frac{b}{2}{\rm sinh}x\right)\right],\\
\psi^-_{(1,2)}&=&\left({\rm cosh}x\right)^{-3/2}\left[
\sin\left(\frac{b}{2}{\rm
sinh}x\right)+\frac{1}{b}(1\mp\sqrt{b^{2}+1}){\rm sinh}x\
\!\cos\left(\frac{b}{2}{\rm sinh}x\right)\right]
\end{eqnarray}
Since we have the exact form of these states, we can calculate the
Wronskians between them exactly. Here, we have
\begin{equation}
\left.W[\psi^{+}_{1},\psi^{+}_{2}]\right|_{x\rightarrow\pm\infty}=
\left.W[\psi^{-}_{1},\psi^{-}_{2}]\right|_{x\rightarrow\pm\infty}=0.
\end{equation}
The positive and the negative parity states separately satisfy the
self-adjoint extension requirement. One can also calculate the
Wronskians between the states in the two sectors,
\begin{eqnarray}
\left.W[\psi^{+}_{1},\psi^{-}_{1}]\right|_{x\rightarrow\pm\infty}
&=&-\frac{1}{2b}\left(b^2+2-2\sqrt{b^2+1}\right),\\
\left.W[\psi^{+}_{1},\psi^{-}_{2}]\right|_{x\rightarrow\pm\infty}
&=&\frac{b}{2},\\
\left.W[\psi^{+}_{2},\psi^{-}_{1}]\right|_{x\rightarrow\pm\infty}
&=&\frac{b}{2},\\
\left.W[\psi^{+}_{2},\psi^{-}_{2}]\right|_{x\rightarrow\pm\infty}
&=&-\frac{1}{2b}\left(b^2+2+2\sqrt{b^2+1}\right).
\end{eqnarray}
Therefore, the two parity sectors together satisfy the
self-adjoint extension requirement that the Wronskian between any
two states has the same limit as $x\rightarrow\pm\infty$.

Finally, there is one more very interesting example with the
potential
\begin{equation}
V(x)=-A_{1}\cosh^{2\nu}x-\frac{\nu}{2}\left(\frac{\nu}{2}+1\right)
\sinh^{2}x.
\end{equation}
This model is discussed recently by Koley and Kar \cite{koley1} in
relation to the localization of fermion fields on branes in a
higher dimensional bulk spacetime \cite{koley2}. The authors found
that for any value of $\nu$, $\nu>0$, one can obtain a pair of
states with the exact form,
\begin{eqnarray}
\psi^+&=&\frac{1}{({\rm cosh}x)^{\nu/2}}\cos\left[\sqrt{A_{1}}\int
\left({\rm cosh}x\right)^{\nu}dx\right],\nonumber\\
\psi^-&=&\frac{1}{({\rm cosh}x)^{\nu/2}}\sin\left[\sqrt{A_{1}}\int
\left({\rm cosh}x\right)^{\nu}dx\right],\label{pair}
\end{eqnarray}
with energy $E=-\nu^{2}/4$. For $\nu=1$ ($A_{1}=b^2/4$), this
model is the same as the quasi-exactly solvable model in
Eq.~(\ref{QES-potential}) above with $n=1$, and this pair of
states is just the parity states constructed from the exactly
solvable total transmission mode. For general values of $\nu$, it
is not hard to see that the pair of states in Eq.~(\ref{pair}) is
also the positive and negative parity states constructed from
total transmission modes with the same energy. It would be
interesting to see if the condition for total transmission
coincides with the quantization rule of the self-adjoint extension
requirement for general $\nu$ in this model.

\section{Summary}

In this paper we have studied the self-adjoint extension procedure
of the Hamiltonian operator with symmetric potentials. These
potentials are unbounded from below but have the property that the
flight time of a classical particle to get to infinity from some
finite reference point, i.e. Eq.~(\ref{T}), is finite.

In this procedure, one makes the requirement that the Wronskian of
any two states approaches the same limit as
$x\rightarrow\pm\infty$.  The Hamiltonian operator can be shown to
be hermitian because the boundary terms cancel. Here the Hilbert
space consists of a discrete set of bound energy eigenstates with
positive and negative parities. Moreover, these bound states exist
both above and below the peak of the potential. This Hilbert space
is characterized by two parameters, the energies of the reference
states in each of the parity sectors. Since the Wronskian is not
required to vanish at infinity, the energy eigenstates could be
degenerate for the total transmission modes.

An interesting special case occurs when the Wronskian between any
two states vanishes as $x\rightarrow\pm\infty$. The Hilbert space
of which the Hamiltonian operator is hermitian again consists of a
discrete set of bound energy eigenstates with positive and
negative parity. This family of states is characterized by only
one parameter, for example, the energy of the reference state in
the spectrum.

\begin{acknowledgments}
This work was supported in part by the National Science Council of
the Republic of China under the Grants NSC 96-2112-M-032-006-MY3
(H.T.C.) and NSC 96-2112-M-032-007-MY3 (C.L.H.). The authors would
also like to thank the National Center for Theoretical Sciences
for partial support.
\end{acknowledgments}

\end{document}